\documentstyle[preprint,aps,epsf,floats]{revtex}    % PREPRINT STYLE
\def\lsim{\mathrel{\raise.2ex\hbox{$<$}\hskip-.8em\lower.9ex\hbox{$\sim$}}}
\def\gsim{\mathrel{\raise.2ex\hbox{$>$}\hskip-.8em\lower.9ex\hbox{$\sim$}}}
\begin{document}
\tighten
%\everymath={\displaystyle}
%
%%%%%%%%%%%%%%%%%%%%% TITLE PAGE %%%%%%%%%%%%%%%%%%%%%%%%%%%%%%%%%%%%%%%%%%%%%%
%
\preprint{
\font\fortssbx=cmssbx10 scaled \magstep2
\hbox to \hsize{
%\special{psfile=uwlogo.ps
%hscale=8000 vscale=8000
%hoffset=-12 voffset=-2}
%\hskip.5in \raise.1in\hbox{\fortssbx University of Wisconsin - Madison}
\hfill$\vtop{   \hbox{\bf TTP95-42}
                \hbox{\bf MADPH-95-916}
%                \hbox{\bf hep-ph/9511???\\}
                \hbox{November 1995}}$ }
}

\title{Dijet Production at HERA in Next-to-Leading Order}
\author{Erwin Mirkes$^a$ and Dieter Zeppenfeld$^b$\\[3mm]}
\address{$^a$Institut f\"ur Theoretische Teilchenphysik, 
         Universit\"at Karlsruhe, D-76128 Karlsruhe, Germany\\[2mm]}
\address{
$^b$Department of Physics, University of Wisconsin, Madison, WI 53706, USA}
\maketitle
\begin{abstract}
Two-jet cross sections in deep inelastic scattering at HERA are calculated in 
next-to-leading order. The QCD corrections are implemented in a new
$ep\rightarrow n$ jets event generator, MEPJET, which allows to analyze 
arbitrary jet definition schemes and general cuts in terms of parton 
4-momenta. First results are presented for the JADE, the cone and the $k_T$
schemes. For the $W$-scheme, disagreement with previous results and large 
radiative corrections and recombination scheme dependencies are traced to a 
common origin.
\end{abstract}
%
%\pacs{PACS numbers: 13.35.+s, 11.30.Rd, 12.40.Vv}
%
\newpage
%
%%%%%%%%%%%%%%%%%% MAIN TEXT %%%%%%%%%%%%%%%%%%%%%%%%%%%%%%%%%%%%%%%%%%%%%%%
%
\narrowtext

Deep inelastic scattering (DIS) at HERA is a copious source of 
multi-jet events. Typical two-jet cross sections\footnote{In the following 
the jet due to the beam remnant is not included in the number of jets.} 
are in the 100~pb to few nb range and thus provide sufficiently high 
statistics for precision QCD tests~\cite{H1,ZEUS}.  
Topics to be studied include the determination of 
$\alpha_s(\mu_R^2)$ over a wide range of scales, measurement of the 
gluon structure function (via $\gamma g\to q\bar q$), and the study of 
internal jet structure. Clearly, next-to-leading order (NLO) QCD corrections
are mandatory on the theoretical side. The dijet cross section, for example,
is proportional to $\alpha_s(\mu_R)$ at leading order (LO), thus suggesting 
a direct measurement of the strong coupling constant. However, the
LO calculation leaves the renormalization scale $\mu_R$ undetermined.
The NLO corrections substantially reduce the renormalization and 
factorization scale dependencies which are present in the LO calculations
and thus reliable cross section predictions in terms of $\alpha_s(m_Z)$
are made possible. NLO corrections in jet physics imply that 
a jet (in a given jet definition scheme) may consist of two partons.
Thus first sensitivity to the internal jet structure is obtained, 
like dependence on the cone size or on recombination prescriptions.
This will be of particular importance in the subsequent discussion.

In this Letter, we investigate NLO QCD corrections to two-jet production 
in various jet algorithms at HERA energies. Previous 
calculations~\cite{wscheme1,wscheme2,disjet} were limited to a JADE type 
algorithm, called $W$-scheme below. In addition, approximations were made 
to the matrix elements which, as we will show, are not valid in large regions
of phase space. Our calculation uses the $s_{min}$ technique of Giele and 
Glover~\cite{giele1}. This technique considerably simplifies the structure 
of NLO QCD corrections to hadronic processes and has already been applied 
to the calculation of NLO jet cross sections at LEP and the Tevatron
\cite{giele1,giele2}. The calculation is based on a QED-type factorization 
of soft gluon singularities \cite{giele1} and the use of universal 
``crossing functions'' \cite{giele2}. This allows for a modular approach to 
NLO calculations. Our calculation of dijet production in DIS is implemented 
as a full NLO Monte Carlo program, MEPJET. The essential benefit of the 
Monte Carlo approach~\cite{jim} is that all hard phase space integrals are 
performed numerically. This makes the implementation of arbitrary cuts and 
jet algorithms a relatively easy task.

Let us briefly discuss the essential features of our calculation. 
Further details will be given in a subsequent publication.
Deep inelastic electron proton scattering with several partons
in the final state,
\begin{equation}
e^-(l) + p(P) \rightarrow  e^-(l^\prime)+
\mbox{proton remnant}(p_r) +
\mbox{parton} \,\,1 (p_1) +
\ldots
+\mbox{parton}\,\, n (p_n)
\label{eq2}
\end{equation}
proceeds via the exchange of an
intermediate vector boson $V=\gamma, Z$. In the following,
$Z$-exchange will be neglected. We denote the %$\gamma^{\ast}$-momentum
momentum of the virtual photon, $\gamma^{\ast}$, by $q=l-l^\prime$, 
its absolute square by $Q^2$, the square of the final hadronic mass by 
$W^2=(P+q)^2$ and use the standard scaling variables 
$x={Q^2}/({2P\cdot q})$ and $y={P\cdot q}/{P\cdot l}$.
The general structure of the $n$-jet cross section in DIS is given by
\begin{equation}
d\sigma^{had}[n-\mbox{jet}] =
\sum_a \int d\eta \,\,f_a(\eta,\mu_F^2)\,\,\, d\hat{\sigma}^a(p=\eta P, 
\alpha_s(\mu_R^2), \mu_R^2, \mu_F^2)
\end{equation}
where the sum runs over incident partons $a=q,\bar{q},g$ which carry 
a fraction $\eta$ of the proton momentum.
$\hat{\sigma}^a$ denotes the partonic cross section from which collinear 
initial state singularities have been factorized out at a scale $\mu_F$ and 
implicitly included in the scale dependent parton densities 
$f_a(\eta,\mu_F^2)$. In Born approximation, the subprocesses  
$\gamma^\ast +q \rightarrow q + g$, 
$\gamma^\ast +\bar q \rightarrow \bar q + g$,  
and
$\gamma^\ast+g \rightarrow q + \bar{q}$  
contribute to the two-jet cross section. 
At ${\cal O}(\alpha_{s}^{2})$ the real emission corrections involve   
$\gamma^\ast+q \rightarrow q + g + g,\,
 \gamma^\ast+q \rightarrow q + \bar{q} + q,\,
 \gamma^\ast+g \rightarrow q + \bar{q} + g\,$
and analogous anti-quark initiated processes. The corresponding 
cross sections are calculated by 
numerically evaluating the tree level helicity amplitudes as given in 
Ref.~\cite{HZ}. Similarly, the finite parts of the one-loop amplitudes 
are obtained by crossing the one-loop results for
$e^+e^- \rightarrow q\bar{q}g$  from Ref.~\cite{giele1}. 
The numerical evaluation of helicity amplitudes has 
the advantage that the full spin structure is kept and, therefore, the 
MEPJET program allows for the calculation of all possible jet-jet and 
jet lepton correlations in NLO. In addition the NLO corrections 
for polarized electron on polarized proton scattering~\cite{ziegler}  
become available.

In the one-loop amplitudes the ultraviolet divergencies are removed by
$\overline{MS}$ renormalization which introduces a dependence on the 
renormalization scale $\mu_R$. 
Infrared as well as collinear divergencies associated
with the final state partons are cancelled against corresponding 
divergencies of the one-loop contributions (see below).
The remaining collinear initial state divergencies are factorized into 
the bare parton densities introducing a dependence on the factorization 
scale $\mu_F$. In order to handle these singularities we follow 
Ref.~\cite{giele2} and use the technique of universal ``crossing
functions''. Starting from the results of the NLO calculation with all 
partons in the final state, {\it i.e.} $e^+e^-\to n+1$~jets, where no such 
singularities occur after adding real and virtual contributions, the 
``crossing functions'' contain the convolutions of the parton
distribution functions with Altarelli-Parisi kernels and in addition
take into account the crossing of a final state cluster to the intitial 
state, within the $s_{min}$ cone as defined below. 
Note that all ``plus'' prescriptions are absorbed into the numerical
evaluation of these crossing functions.

The 3-parton final states need to be integrated over the entire phase space,
including the unresolved regions, where only two jets 
are reconstructed according to a given jet definition scheme.  In order
to isolate the infrared as well as collinear divergencies associated with
these unresolved regions the resolution parameter $s_{min}$ is introduced.
Soft and collinear approximations are used in the region where at least one 
pair of partons, including initial ones, has $s_{ij}=2p_i\cdot p_j<s_{min}$
and the soft and/or collinear final state parton is integrated over 
analytically. Adding this soft+collinear part to the virtual contributions 
gives a finite result for, effectively, 2-parton final states. In general 
this 2-parton contribution is negative and
grows logarithmically in magnitude as $s_{min}$ is decreased. This logarithmic 
growth is exactly cancelled by the increase in the 3 parton cross 
section, once $s_{min}$ is small enough for the approximations to be valid.
The integration over the 3-parton phase space with $s_{ij}>s_{min}$
is done by Monte-Carlo techniques (without using
any approximations). Since, at each phase space point, the parton 4-momenta 
are available, the program is flexible enough to implement 
any jet definition algorithms or to impose arbitrary
kinematical resolution and acceptance cuts on the final state particles.
This is an essential advantage over existing programs such as 
DISJET~\cite{disjet}.

\setlength{\unitlength}{0.7mm}
\begin{figure}[hbt]               \vspace*{-4cm}
\begin{picture}(150,165)(-30,1)
\mbox{\epsfxsize10.0cm\epsffile[70 250 480 550]{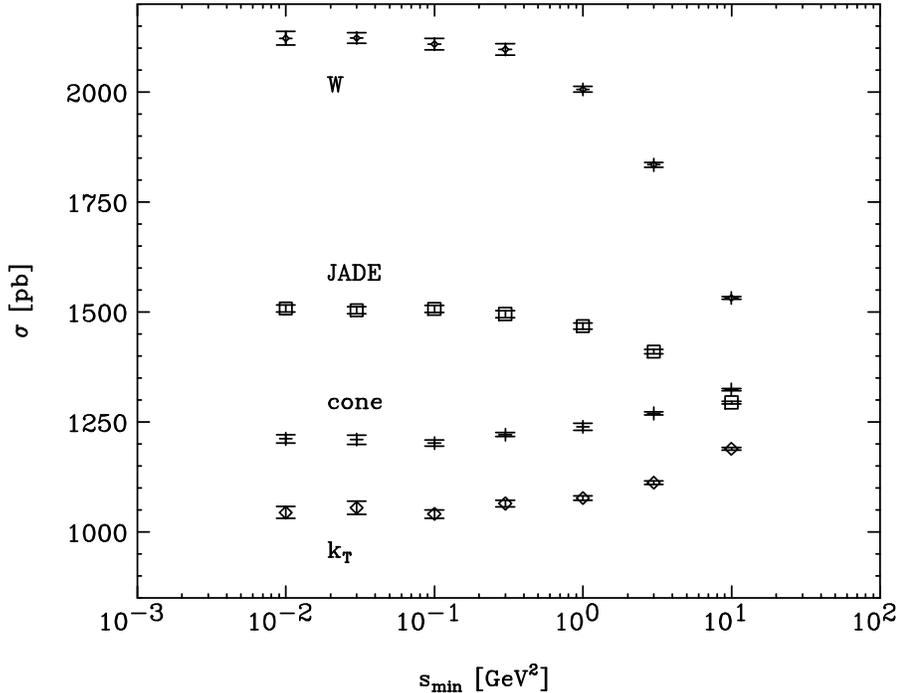}}
\end{picture}
\vspace*{4cm}
\caption{
Dependence of the inclusive two-jet cross section
in the $k_T$, cone, JADE, and the
$W$-scheme on $s_{min}$, the two-parton resolution parameter. 
Partons are recombined in the $E$-scheme. 
Error bars represent statistical errors of the Monte Carlo program. 
For the fairly soft jet definition criteria described in the text, 
$s_{min}$ independence is achieved for $s_{min}\lsim 0.1$~GeV$^2$. 
}\label{fig1}
\end{figure}

A powerful test of the numerical program is the $s_{min}$ independence of 
the final result. Fig.~1 shows the inclusive dijet cross section as a 
function of $s_{min}$ for the jet algorithms to be defined below. 
As mentioned before, $s_{min}$ is an arbitrary theoretical parameter 
and any measurable quantity should not depend on it. One observes
that for values smaller than 0.1~GeV$^2$ the results are indeed  
independent of $s_{min}$. The strong $s_{min}$ dependence of the NLO 
cross sections for larger values shows that the soft 
and collinear approximations used in the phase space region $s_{ij}<s_{min}$
are no longer valid, {\it i.e.} terms of ${\cal{O}}(s_{min})$ and
${\cal{O}}(s_{min}\ln s_{min})$ become important.
In general, one wants to choose $s_{min}$ as large as possible to avoid large
cancellations between the virtual+collinear+soft part ($s_{ij}<s_{min}$)
and the hard part of the phase space  ($s_{ij}>s_{min}$).
Note that factor 10 cancellations occur between the effective 2-parton and 
3-parton final states at the lowest $s_{min}$ values in Fig.~1 
and hence very high 
Monte Carlo statistics is required for these points. $s_{min}$ independence 
is achieved at and below $s_{min}=0.1$~GeV$^2$ and we choose this value for 
our further studies.

%All integrations over the hard phase space are done by Monte-Carlo
%techniques, which provides great flexibility to impose arbitrary
%cuts and to investigate different jet definition schemes. 
For these numerical
studies, the standard set of parton distribution functions is MRS set 
D-' \cite{mrs}.
We employ the two loop formula for the strong coupling constant
%
%
%\begin{equation}
%\alpha_{s\,\overline{MS}}(\mu_R^2)= 
%\frac{12\pi}{(33-2n_{f})\ln(\mu_R^2/\Lambda^{2})}
%\left[1-\frac{6(153-19n_{f})}{(33-2n_{f})^{2}}
%\frac{\ln\ln(\mu_R^2/\Lambda^{2})}{\ln(\mu_R^2/\Lambda^{2})}\right]
%\label{asnlo}
%\end{equation}
%
%
with %${\Lambda_{\overline{MS}}}={\Lambda_{\overline{MS}}^{(4)}}=230$ MeV, 
${\Lambda_{\overline{MS}}^{(4)}}=230$ MeV, 
which is the value from the parton distribution functions.
The value of $\alpha_s$ is matched at the thresholds $\mu_R=m_q$ and the
number of flavors is fixed to $n_f=5$ throughout, {\it i.e.} gluons are 
allowed to split into five flavors of massless quarks.
Unless stated otherwise, the renormalization scale  and
the factorization scale  are set to
$\mu_R=\mu_F= 1/2\,\sum_i \,p_T^B(i)$, where $p_T^B(i)$ denotes the magnitude
of the transverse momentum of parton $i$ in the Breit frame.
A running QED fine structure constant $\alpha(Q^2)$ is used.
The lepton and hadron beam energies are 27.5 and 820 GeV, respectively.
A  minimal set of kinematical cuts is imposed on the initial virtual photon 
and on the final state electron and jets. We require 
40~GeV$^2<Q^2<2500$ GeV$^2$,
$0.04 < y < 1$, an energy cut of $E(e^\prime)>10$~GeV on the scattered 
electron, and a cut on the pseudo-rapidity $\eta=-\ln\tan(\theta/2)$
of the scattered lepton and jets of $|\eta|<3.5$. In addition jets 
must have transverse momenta of at least 2~GeV in both the lab and the 
Breit frame.

Within these general cuts four different jet definition schemes are 
considered for which we have chosen parameters such as to give similar
LO cross sections (see Table~I). In the $W$-scheme the invariant mass squared 
$s_{ij}=(p_i+p_j)^2$ is calculated for each pair of final state 
particles (including the proton remnant).
If the pair with the smallest invariant mass squared is below $y_{cut}W^2$,
the pair is clustered according to a recombination scheme. 
Unless stated otherwise, and for all
jet algorithms, we use the $E$-scheme to recombine partons, i.e.
the cluster momentum is taken as $p_i+p_j$. This process 
is repeated until all invariant masses are above $y_{cut} W^2$.        
The  resolution parameter $y_{cut}$ is fixed to $0.02$.

The experimental analyses in \cite{H1,ZEUS} are  based 
on a variant of the $W$-scheme, the ``JADE'' algorithm. It is
obtained from the $W$-scheme by replacing the invariant
definition $s_{ij}=(p_i+p_j)^2$ by 
$M_{ij}^2=2E_iE_j(1-\cos\theta_{ij})$, where all quantities are defined 
in the laboratory frame. Neglecting the explicit mass terms $p_i^2$ 
and $p_j^2$ in the definition of $M_{ij}^2$ causes substantial
differences in jet cross sections between the $W$ and the JADE scheme.

In the cone algorithm (which is defined in the laboratory frame) the distance 
$\Delta R=\sqrt{(\Delta\eta)^2+(\Delta\phi)^2}$ between two partons 
decides whether they should be recombined to a single jet. Here the variables 
are the pseudo-rapidity $\eta$ and the azimuthal angle $\phi$. We 
recombine partons with $\Delta R<1$.
Furthermore, a cut on the jet transverse momenta of $p_T(j)>5$~GeV in the lab 
frame is imposed in addition to the 2 GeV Breit frame cut.

For the $k_T$ algorithm (which is implemented in the Breit frame), 
we follow the description introduced
in Ref.~\cite{kt}. The hard scattering scale $E_T^2$ is
fixed to 40 GeV$^2$ and $y_{cut}=1$ is the resolution parameter 
for resolving the macro-jets. In addition, jets are required to have a minimal 
transverse momentum of 5 GeV in the Breit frame.

\begin{table}[t]
\caption{Two-jet cross sections in DIS at HERA. Results are given at LO and
NLO for the four jet definition schemes and acceptance cuts described in 
the text.
}\label{table1}
\vspace{2mm}
\begin{tabular}{lccc}
        \hspace{2.3cm}
     &  \mbox{two-jet }
     &  \mbox{two-jet exclusive}
     &  \mbox{two-jet inclusive} \\
     &  \mbox{LO}
     &  \mbox{NLO}
     &  \mbox{NLO}\\
\hline\\[-3mm]
\mbox{cone} & 1107~pb & 1047~pb & 1203~pb  \\
$k_T$       & 1067 pb & 946 pb  & 1038 pb  \\
$W$         & 1020 pb & 2061 pb & 2082 pb  \\
\mbox{JADE} & 1020 pb & 1473 pb & 1507 pb  \\
\end{tabular}
\end{table}

With these parameters, one obtains the two-jet cross sections of 
Table~\ref{table1}. While the higher order corrections in the cone
and $k_T$ schemes are small, very large corrections appear in the $W$-scheme. 
In addition, the large effective $K$-factor 
(defined as $K=\sigma_{NLO}/\sigma_{LO}$) 
of 2.04 (2.02) for the two-jet inclusive
(exclusive) cross section in the $W$-scheme depends strongly 
on the recombination scheme which is used in the
clustering algorithm (see below).

\setlength{\unitlength}{0.7mm}
\begin{figure}[hbt]               \vspace*{-2cm}
\begin{picture}(150,165)(-30,1)
\mbox{\epsfxsize10.0cm\epsffile[78 222 480 650]{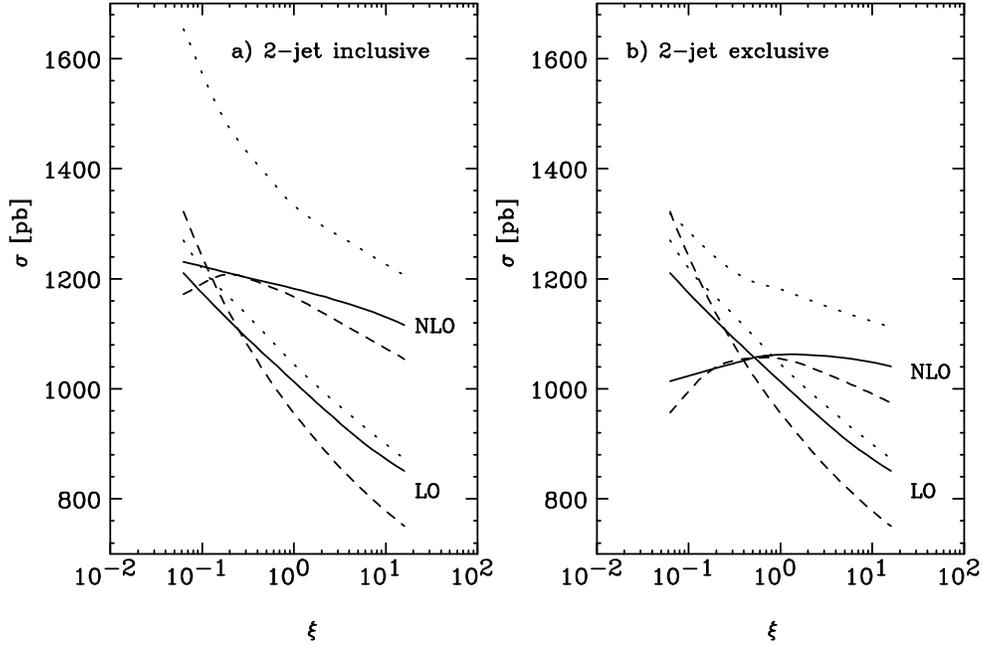}}
\end{picture}
\caption{
Dependence of a) the two-jet inclusive and b) the two-jet exclusive cross 
section in the cone scheme 
on the renormalization and factorization scale factor $\xi$.
The solid curves are for $\mu_R^2=\mu_F^2=\xi\;(\sum_i\;p_T^B(i))^2$, while for
the dashed curves only $\xi_R=\xi$ is varied but $\xi_F=1/4$ is fixed.
Choosing the photon virtuality as the basic scale yields the dotted curves,
which correspond to $\mu_R^2=\mu_F^2=\xi\;Q^2$. Results are shown 
for the LO (lower curves) and NLO calculations. 
}\label{fig2}
\end{figure}

A good measure of the improvement at NLO is provided by the residual scale 
dependence of the cross section. We have considered scales related to the 
scalar sum of the parton transverse momenta in the Breit frame,
$\sum_i \,p_T^B(i)$, and the virtuality $Q^2$ of the incident photon.
In Fig.~\ref{fig2} the dependence of the two-jet 
cross section, in the cone scheme, on the renormalization and 
factorization scale factors $\xi_R$ and $\xi_F$ is shown. For scales related 
to $\sum_i \,p_T^B(i)$ they are defined via
\begin{equation}
 \mu_R^2 = \xi_R\;(\sum_i \,p_T^B(i))^2\, ,
\hspace{1cm}
 \mu_F^2 = \xi_F\;(\sum_i \,p_T^B(i))^2\,.
\label{xidef}
\end{equation}
For the two-jet inclusive cross section of Fig.~\ref{fig2}a,
the LO variation by a factor 1.43 is reduced to a 10\%
variation at NLO when both scales are varied simultaneously 
over the plotted range (solid curves).
However, neither the LO nor the NLO curves show an extremum. 
The uncertainty from the variation of both scales  for the NLO two-jet
exclusive cross section in Fig.~2b (solid curves) is reduced to 5\%.
Furthermore, the two-jet exclusive cross section now has a 
maximum and is equal to the LO cross section for $\xi=0.5$.
Also shown is the $\xi=\xi_R$ dependence of LO and NLO cross
sections at fixed $\xi_F=1/4$ (dashed curves). In this case 
a maximum appears in the NLO inclusive and exclusive cross sections. 
However, the scale variation is stronger than in the $\xi=\xi_R=\xi_F$ case.

An alternative scale choice might be $\mu_R^2=\mu_F^2=\xi\,Q^2$.
The resulting $\xi$ dependence is shown as the dotted lines for
both the LO and NLO calculations.
At LO the two scale choices give qualitatively similar results.
However, with $\mu_R^2=\mu_F^2=\xi\,Q^2$, the scale dependence 
does not markedly  improve at NLO.
% within the kinematical range considered here. 
In addition a sizable $K$-factor is found, with $K>1$ for small values
of $Q^2$ and $K<1$ for very hard incident photons. We therefore use 
the jet transverse momenta in the Breit frame to set the scale
and fix $\xi_R=\xi_F=1/4$ in Eq.~(\ref{xidef}).
A further discussion of the scale dependence of dijet cross sections at HERA
can be found in Ref.~\cite{ingelman}.

The effective $K$-factors close to unity which are found in the cone and 
$k_T$ schemes could, in principle, be a coincidence arising from compensating 
effects in different phase space regions. It is important, therefore, to also 
compare LO and NLO distributions, in particular for those variables which
define the acceptance region. 

\setlength{\unitlength}{0.7mm}
\begin{figure}[hbt]               \vspace*{-2cm}
\begin{picture}(150,165)(-30,1)
\mbox{\epsfxsize10.0cm\epsffile[78 222 480 650]{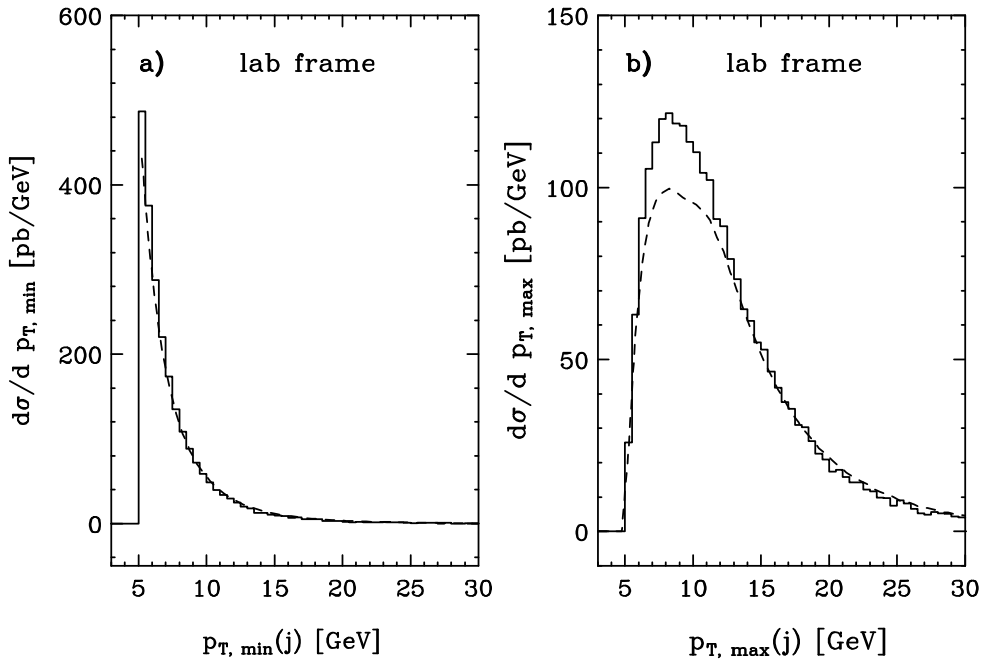}}
\end{picture}
\caption{
Transverse momentum distribution in the lab frame
for the jet with (a) minimal and (b) maximal transverse momentum.
Results are shown for the two-jet inclusive cross
section in the cone scheme in leading (dashed curves)
and next-to-leading  order (histograms).
}
\end{figure}

The transverse momentum distributions of the 
softest and the hardest jet in the laboratory frame are shown in Fig.~3, for
the cone scheme. 
The fairly small NLO corrections allow for reliable theoretical predictions.
In general the largest radiative corrections are observed at small jet $p_T$,
as evidenced by the shape change in the 
$p_{T,max}$ distribution of Fig.~3b. The 
predictions are therefore expected to become more reliable 
for higher jet transverse momenta.
A potential problem is the very steep $p_{T,\,min}^{lab}$ distribution,
which, via the cut at 5~GeV, introduces a strong sensitivity to the 
correct matching of the parton $p_T$ and the measured jet $p_T$.
However, this is a general problem for all jet algorithms, {\it i.e.}
the jet rate falls very rapidly as the required energy scale of the jets 
is increased.

\setlength{\unitlength}{0.7mm}
\begin{figure}[t]               \vspace*{-2cm}
\begin{picture}(150,165)(-30,1)
\mbox{\epsfxsize10.0cm\epsffile[78 222 480 650]{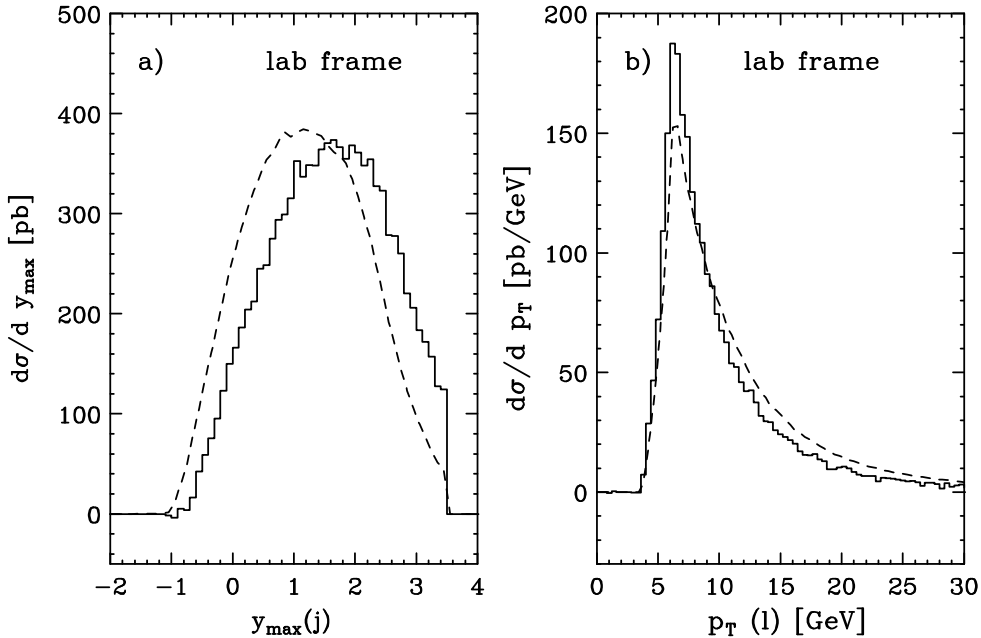}}
\end{picture}
\caption{
Rapidity distribution of the most forward jet (a) and 
transverse momentum distribution of the scattered electron (b)
in the lab frame. Results are shown for the  $k_T$ scheme in leading 
(dashed curves) and next-to-leading order (histograms) for the 
two-jet inclusive cross section.
}
\end{figure}

A more critical case is shown in Fig.~4 where, for jets defined in the 
$k_T$-scheme, the jet rapidity and the electron transverse momentum in the 
lab frame are shown. At NLO jets are produced somewhat more 
forward (in the proton direction) than at LO, see Fig.~4a.
Hence, the rapidity cut at $|\eta_j|=3.5$ has a stronger effect in NLO,
which partially explains the relatively low 
$K$-factor of 0.97 in the $k_T$-scheme.

Another observable which exhibits rather large NLO corrections is the 
electron transverse momentum distribution in Fig.~4b. The electron $p_T$
becomes considerably softer at NLO, with an effective $K$-factor above 
unity at small $p_T(\ell)$ and $K<1$ in the high transverse momentum 
region. In view of these shape changes the overall small change at NLO
has to be considered a coincidence, tied to the choice of $p_T(\ell)$ 
range. Since the electron transverse 
momentum and the $Q^2$ of the event are very closely related, a similar 
change in the size of radiative corrections is obtained by choosing 
different $Q^2$ bins. Very similar effects on the $y_{max}(j)$ and 
$p_T(l)$ distributions are also observed in the other jet definition schemes.
As a result, a judicious choice of phase space region could generate 
very large or small $K$-factors which would indicate that, in these phase 
space regions, even the NLO calculation is fraught with large uncertainties.
To avoid such potential problems,
one should investigate the effect of the higher order corrections
on those variables which are used to define kinematical cuts.

A somewhat disturbing result of our calculation is the surprisingly large 
effective $K$-factor in the $W$-scheme, in particular since we disagree here
with previous calculations
\footnote{About 10\% are explained by differences 
in the virtual results for the ``longitudinal projection'' in the first 
paper of \cite{wscheme2}.
This has been corrected in the latest version of \cite{disjet}.}
\cite{wscheme2,disjet}.
The DISJET program\cite{disjet}, for example, gives a $K$-factor very 
close to unity for a phase space region which is very similar\footnote{
The DISJET program does not allow to impose exactly the same kinematical
cuts as discussed before for the $W$-scheme; however, this is
not relevant for the argument.} 
to the one considered above. The main difference to these
earlier calculations is the treatment of partons which are recombined into 
a jet. Previously, collinear and soft approximations were made for all
partons which are coalesced into a single  jet, {\it i.e.} for any pair 
of final state particles (including the remnant) with 
$s_{ij} < y_{cut}\;W^2$, neglecting terms of order $y_{cut}W^2$.
%The integration over the final state singular regions, for example,
% was performed in the rest frame of the 
%``unresolved parton pair'' %(defined by $s_{ij} < y_{cut}\;W^2$)
%up to  ${\cal{O}}(y_{cut}^0)$ accuracy,
%{\it i.e.} terms proportional to $y_{cut}$ were neglected. 
In our calculation, similar 
\footnote{Our approximations in the soft and collinear region
are stronger than the approximations made in \cite{wscheme2,disjet}
where, effectively, also terms proportional to $s_{ij}\ln s_{ij}$ 
are kept.  These latter terms are neglected here.} 
approximations are only made in a much smaller region, $s_{ij}< s_{min}$ 
for any pair of initial and final state partons, and we are able to study the
small $s_{min}$ limit. 
Effectively, the approximations in Refs.~\cite{wscheme2,disjet}
correspond to the replacement of a massive two-parton 
system by a single massless jet.
How well justified are these approximations? 

\setlength{\unitlength}{0.7mm}
\begin{figure}[hbt]               \vspace*{-2cm}
\begin{picture}(150,165)(-30,1)
\mbox{\epsfxsize10.0cm\epsffile[78 222 480 650]{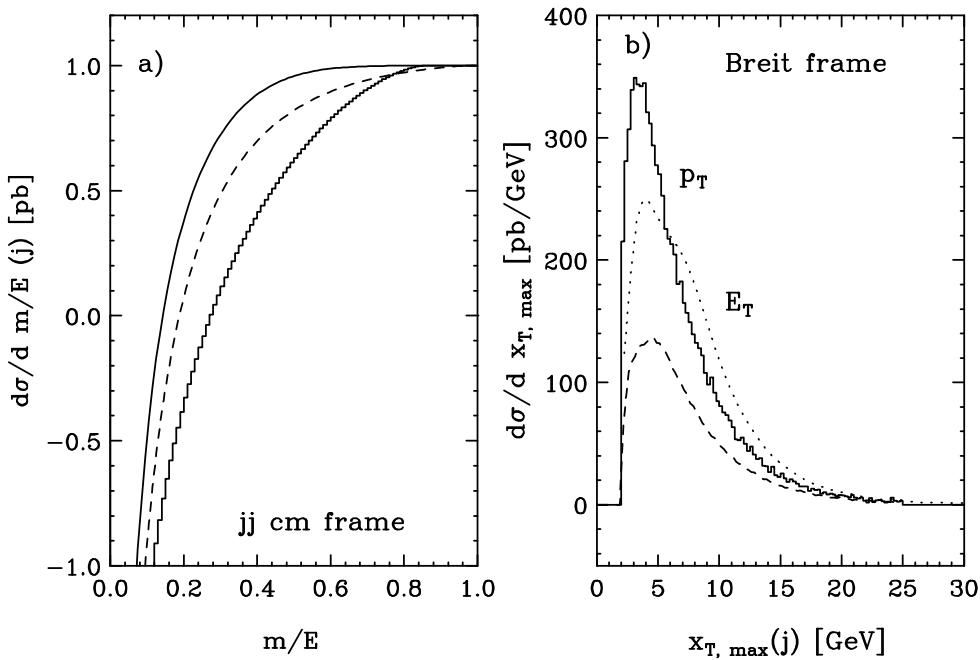}}
\end{picture}
\caption{
Single jet mass effects at next-to-leading order. 
(a) Fraction of events in the cone scheme (solid curve), 
$k_T$ scheme (dashed curve), and $W$-scheme (histogram) with all jet 
mass to energy ratios below $m/E$, where $E$ is the corresponding 
jet's energy in the parton center of mass frame. Negative values at small 
$m/E$ are due to virtual contributions at $m/E=0$.
(b) Next-to-leading order transverse momentum ($x_T=p_T$, solid histogram) 
   and transverse-energy  distribution ($x_T=E_T$, dotted curve)
   for the jet with largest $p_T$ and $E_T$ in the Breit frame, for
   the $W$-scheme. The dashed curve shows the leading order result
   where both distributions are identical.
}
\end{figure}

In Fig.~5a the fraction of events is shown with at least one jet being more 
massive than  $m/E$. Here $m$ is the invariant mass and $E$ the energy of 
the most massive of the jets in the parton center of mass frame.
%jet-jet center of mass frame.
In LO $m/E\equiv 0$ since we are always using massless partons. At NLO
the median $m/E$ is 0.44 in the $W$-scheme
({\it i.e.} 50\% of the NLO cross section lies in the range
$m/E>0.44$), while substantially smaller values,
0.30 and 0.22 are found in the $k_T$ and cone schemes. The large single jet
masses (compared to their energy) imply that collinear or soft approximations 
for the partons inside a NLO jet are not generally permissable. Indeed the slow
approach to the asymptotic region visible in Fig.~1 shows that only for 
two parton invariant masses squared $s_{ij}\lsim 0.1$~GeV$^2$ are such 
approximations allowed. It is these mass effects which were not treated 
correctly in previous calculations.
%, a restriction which is certainly not obeyed in the 
% calculations onto which the PROJET and DISJET Monte Carlos are based.

The very large median value of $m/E$ in the $W$-scheme implies that at NLO 
we are dealing with very different types of jets than at LO. At NLO at least 
one of the $W$-scheme jets extends over a large solid angle, it is a massive,
slow moving object in the center of mass frame and, hence, very different from
the pencil-like, massless objects called jets at LO. The typically small 
relativistic $\gamma$-factor of these jets has large kinematic effects. 
For example the difference between transverse energy and transverse momentum 
distributions of the jets,  which are shown in Fig.~5b, becomes quite 
pronounced in the $W$-scheme, an effect which is much smaller in the $k_T$ and 
cone schemes.

The fact that very different kinematical regions are populated 
by the LO and NLO jets implies that radiative corrections may be very large
as is indeed indicated by the large effective $K$-factor in the $W$-scheme.
The large corrections in turn imply that the predicted jet cross sections have
large theoretical uncertainties and should not be used for precise QCD tests.
Another measure of these uncertainties is the recombination scheme dependence
of the predicted two-jet inclusive cross section. In the $E0$ and $P$-schemes,
where two partons are recombined to a massless jet, $K$ is reduced to
1.41 and 1.29, respectively. This dependence first appears in the NLO 
calculation, where a jet may be composed of two partons. This internal jet
structure, however, is only simulated at tree level and thus the dependence 
of the cross section on the recombination scheme is subject to potentially
large higher order corrections.

These uncertainties are small for jet algorithms with small recombination 
scheme dependence. In the JADE-algorithm the $K$-factor is reduced from 
1.48 in the $E$-scheme to 1.36 and 1.24 in the $E0$ and $P$-schemes.
For the  cone ($k_T$) scheme this recombination scheme
dependence is reduced to the 3\% (10\%) level.
The $K$-factor close to unity as well as this small recombination scheme 
dependence indicate that 
theoretical uncertainties due to higher order
corrections 
are fairly small in the cone and $k_T$ scheme, 
and both schemes appear to be well suited for QCD studies at HERA.

\acknowledgements
D.~Z. would like to thank the CERN theory group, where part of this work 
was completed, for its hospitality.
E.~M. is happy to acknowledge many stimulating conversations
with the ZEUS collaboration and in particular with
B.~Musgrave, S.~Magill, I.~Park, J.~Repond and T.~Trefzger.
This research was supported by the University of Wisconsin 
Research Committee with funds granted by the Wisconsin Alumni Research 
Foundation and by the U.~S.~Department of Energy under Grant 
No.~DE-FG02-95ER40896. The work of E.~M. was supported in part  
by DFG Contract Ku 502/5-1.

%%%%%%%%%%%%%%%%%%%%% REFERENCES %%%%%%%%%%%%%%%%%%%%%%%%%%%%%%%%%%%%%%%%%%%%%%
%

%\end{thebibliography}
%


\begin{references}
%
\bibitem{H1}
H1 Collaboration, T.~Ahmed et. al., 
Phys. Lett. {\bf B346} (1995) 415.

\bibitem{ZEUS}
ZEUS Collaboration, M. Derrick et al.,
DESY-95-182, Oct 1995,
hep-ex/9510001.

\bibitem{wscheme1} 
J.G. K\"orner, E. Mirkes and G. Schuler,
Int. J. Mod. Phys.  {\bf A4} (1989) 1781;\\
T. Brodkorb, J.G. K\"orner, E. Mirkes, and G. Schuler,
Z.~Phys.~ {\bf C44} (1989) 415.

\bibitem{wscheme2} 
T. Brodkorb and J.G. K\"orner, Z.~Phys.~{\bf C54} (1992) 519;\\
T.~Brodkorb and E.~Mirkes, Z.~Phys.~{\bf C66} (1995) 141;\\
D. Graudenz Phys. Lett. {\bf B256} (1992) 518;
            Phys. Rev. {\bf D49} (1994) 3291.

\bibitem{disjet} 
T.~Brodkorb and E.~Mirkes, Madison preprint, MAD/PH/821 (1994).

\bibitem{giele1}
W.~T.~ Giele and E.~W.~N.~Glover,
Phys. Rev. D {\bf 46} (1992) 1980.

\bibitem{giele2}
W.~T.~ Giele, E.~W.~N.~Glover and D.A. Kosower, 
Nucl. Phys. {\bf B403} (1993) 633.

\bibitem{jim} 
H.~Baer, J.~Ohnemus, and J.~F.~Owens,
Phys. Rev.  {\bf D40} (1989) 2844;
Phys. Lett. {\bf B234} (1990) 127;
Phys. Rev.  {\bf D42} (1990) 61.

\bibitem{HZ} 
K.~Hagiwara and D.~Zeppenfeld, Nucl. Phys. {\bf B313} (1989) 560.

\bibitem{ziegler} 
C.~Ziegler  and E.~Mirkes,  Nucl. Phys. {\bf B429} (1994) 93. 

\bibitem{mrs}  
A.D. Martin, W.J. Stirling and R.G. Roberts,
Phys. Lett. {\bf B306} (1993) 145.          

\bibitem{kt}
S.~Catani, Y.L.~Dokshitzer and B.R.~Webber,
Phys. Lett. {\bf B285} (1992) 291.          

\bibitem{ingelman}
G. Ingelman and J. Rathsman,
Z.~Phys.~ {\bf C63} (1994) 589.


\end{references}
\end{document}